\documentstyle[preprint,revtex]{aps}

\begin{document}
\draft

\begin{title}
Spin Gap and Superconductivity in the \\
One-Dimensional \mbox{$t$-$J$}\ Model with Coulomb Repulsion
\end{title}

\author{M. Troyer$^{(a,b,c)}$, H. Tsunetsugu$^{(a,b)}$, T.M.Rice$^{(b)}$ \\
J. Riera $^{(d)}$ and E. Dagotto$^{(d)}$}
\begin{instit}
$^{(a)}$Interdisziplin\"{a}res Projektzentrum
f\"{u}r Supercomputing, \\
Eidgen\"ossische Technische Hochschule, CH-8092 Z\"urich,
Switzerland \\
$^{(b)}$Theoretische Physik, Eidgen\"ossische Technische
Hochschule,\\
CH-8093 Z\"urich, Switzerland\\
$^{(c)}$Centro Svizzero di Calcolo Scientifico, CH-6924 Manno,
Switzerland \\
$^{(d)}$Department of Physics, Center for Materials Research and
Technology, and \\
Supercomputer Computations Research Institute, \\
Florida State University,
Tallahassee, FL 32306, USA. \\
\end{instit}
\receipt{}

%
%
\begin{abstract}
The one-dimensional \mbox{$t$-$J$}\ model with density-density
repulsive interactions is
investigated using exact diagonalization and quantum Monte Carlo methods.
A short-range repulsion pushes phase separation to larger values of
$J/t$, and leads to a widened precursor region in which a spin gap
and strengthened superconducting correlations appear.  The correlation
exponent is calculated.  On the contrary, a
long-range repulsion of $1/r$-form
suppresses superconductivity in the precursor region.
\end{abstract}
\pacs{}

%
%
\section{Introduction}

The \mbox{$t$-$J$}\ model is one of the most actively studied model
Hamiltonians for the cuprate superconductors \cite{anderson,zhang}.
Two essential features of these materials are correctly implemented
in this model, namely the strong on-site Coulomb repulsion at
\mbox{$Cu$}\ sites and strong antiferromagnetic (AF) spin fluctuations
in the ground state.
Although much effort has been devoted to the search for
superconductivity in this model, the parameter values where
superconductivity may appear and the symmetry of the superconducting
order parameter remain to be definitively established.
Two arguments have been given that suggest the presence of
superconducting
correlations in this model.
One proposal is based on the strong AF fluctuations
that occur in the region of small hole doping away from the AF-ordered
state at half-filling \cite{aftheory}.  The second is based on the
occurrence of phase separation at larger values of $J/t$ \cite{phasesep}.
In particular, in several papers it has been
argued that superconductivity may appear
in the precursor region to phase separation.
Only recently, numerical studies have been carried out in this region.
Dagotto and Riera
\cite{dagotto} examined the \mbox{$t$-$J$}\ model at quarter-filling
in one (1D) and two (2D) dimensions.  In particular, they found that adding a
short-range density-density repulsion pushes phase separation to larger values
of $J/t$ and leads to strengthened
signs of superconducting correlations. Those authors provided an
intuitive physical picture of the origin of this effect, based on the
effect of the nearest-neighbor repulsion to form pairs of electrons in
the ground state \cite{kivelson}. The analysis in 2D is similar to what
occurs in 1D. In particular, Dagotto and Riera have recently shown
that a superconducting  state with $d_{x^2-y^2}$ symmetry may exist in
the more realistic case of the 2D
 \mbox{$t$-$J$}\ model near phase separation \cite{dagotto2}.
A recent study of a 1D two band model also shows
that superconducting correlations are enhanced near the phase
separation boundary \cite{subdo}.

Although the parameter region where superconductivity was observed
is not obviously
directly relevant for the \mbox{high-$T_c$}\ cuprates, it is
not excluded that both regions are analytically connected.
Thus, it is important to continue the study
of this phenomenon. This is the purpose of the present paper.
Our calculations are restricted to 1D for simplicity. In this case
a reliable finite-size scaling analysis can be implemented, and exact
diagonalization results can be extrapolated to infinite size.
Another advantage of the 1D problem is
that quantum Monte Carlo (QMC) methods can be used
on large samples down to low temperatures without encountering
the sign problem. Finally, the correlation exponents can be conveniently
calculated using conformal field theory (CFT). Previous
studies \cite{dagotto} have convincingly shown that there are strong
analogies
between 1D and 2D systems, and thus our results may be valid also in
higher dimensions.

The 1D \mbox{$t$-$J$}\ model without repulsive density-density
interactions has been
investigated in detail by Ogata et al. \cite{ogata} using finite size
diagonalization, and by Assaad and W\"urtz \cite{assaad} using QMC
methods.  This 1D \mbox{$t$-$J$}\ model shows three different ground
state behaviors in the plane defined by the electron density,
$\rho$, and the ratio of spin exchange interaction to hopping amplitude,
$J/t$.  First, a Tomonaga-Luttinger liquid (TLL) exists at all
$\rho \ne 1$ for not so large exchange coupling $J/t$.
This phase is characterized
by gapless charge and spin excitations, and by power-law correlation
functions in space and time \cite{solyom}.  Second, there is a
phase-separated region for large $J/t$ at all densities $\rho$.
In this region,
the charge fluctuations are completely suppressed except for the
center-of-mass motion of the electron condensed part, while
the spin excitations are gapless.  Third, there is a
spin gap phase \cite{ogata}, which exists between the previous two regimes
but at small $\rho$-values (at $\rho=1/3$ or larger the spin gap is no longer
observed).  It has gapless
charge excitations, but exponentially decaying spin correlation
functions.  In the regions with gapless charge excitations, power-law
behaviors of correlation functions are determined by a single parameter
\mbox{$K_\rho$}\ \cite{solyom,haldane}.   Superconductivity is the
most dominant correlation in the regime where $\mbox{$K_\rho$} >1$,
for both the TLL and the spin gap phases.  This region is found close
to the phase separation boundary \cite{ogata} and is caused by
an attractive interaction via spin exchange.  In the large $J/t$
region, phase separation overcomes the competition with
superconductivity.  Therefore, at least naively
we may expect a wider superconducting
region if we can suppress phase separation by including other
terms in the Hamiltonian.
This can be easily achieved.
Phase separation can be suppressed by the long-range
part of the Coulomb repulsion between electrons.  This long-range term
is neglected in the Hubbard and \mbox{$t$-$J$}\
models.  However, this term plays a crucial role in and near
the phase separated region.  In this paper we will study the problem of
whether the long-range Coulomb repulsion can enhance the superconductivity
by suppressing phase separation and causing a large precursor region.

Another central issue of this paper is to analyze if this precursor
region is characterized by a gap in the spin excitation spectrum.
As proposed first by Anderson \cite{anderson} and by Kivelson et al.
 \cite{kivelson2} afterwards, a possible mechanism of
the \mbox{high-$T_c$}\ superconductivity is that upon doping of
holes local spin-singlet electron pairs, which would constitute
a spin liquid state in the undoped insulator, start to move coherently,
resulting in an off-diagonal long-range order (ODLRO).  If the
hole motion does not destroy the local character of singlet pairs
completely, the lowest spin excitation is given by a singlet-to-triplet
excitation of a local pair, and thus it will have a finite gap, even though
its absolute value may be reduced by virtual pair-breaking effects due
to charge fluctuations.  Therefore, a finite spin gap is indicative of
the short-range
resonating valence bond (RVB) mechanism of superconductivity.
Following this scenario, the effect of hole doping has been
examined for several models which have an insulating ground state
with a spin gap at half filling \cite{imada,ogata3}, and strong
superconducting correlations are found upon doping.  In this paper,
we will study the models where a spin gap is caused by doping holes
into an antiferromagnetic insulator with gapless spin excitations.
This is consistent with experiments on several \mbox{high-$T_c$}\
materials, particularly in underdoped materials \cite{rice,spingap}.
However, the question of how these local spin-singlet pairs are
stabilized when holes are mobile is not well
understood.
The existence of a spin gap also has a strong relation
with the internal symmetry of superconductivity.  In the TLL phase
(i.e., no spin gap), the nearest-neighbor singlet and triplet
pairing have the same exponent in the superconducting correlation
functions \cite{solyom}.
 Once the spin gap becomes finite by changing coupling constants
in the Hamiltonian or electron concentration, the system is
scaled to another universality class, the Luther-Emery fixed
point in the $g$-ology \cite{solyom,lutheremery}.
Thus, the line where the gap opens
corresponds to the phase boundary
between two different universality classes, the TLL phase
and the spin gap (Luther-Emery) phase, and physical properties
change their long-range behaviors drastically on this line \cite{solyom}.
In the spin gap phase, the triplet and singlet superconductivity
are no longer
degenerate and the triplet superconductivity is suppressed relative
to the singlet one, since making a triplet pair costs a finite energy.

In this paper, we will study the effects of long-range Coulomb repulsion
on superconducting correlations and the spin gap, using three modifications
of the \mbox{$t$-$J$}\ model with different interaction ranges.
The first one is the \mbox{$t$-$J$}\ model plus nearest-neighbor
repulsions \cite{dagotto,kivelson,didier},
\begin{eqnarray} \label{eq:htjv}
    H_{t-J-V}&=&-t\sum_{i,\sigma}
        \left({\cal P}c_{i,\sigma}^{\dag}c_{i+1,\sigma}{\cal P}+h.c.\right)
          \nonumber \\
	&&+J \sum_{i}\left({\bf S}_i \cdot {\bf S}_{i+1}
	-{1\over4}n_in_{i+1}\right)+ V\sum_{i}n_in_{i+1} \; ,
\end{eqnarray}
where $c_{i,\sigma}^{\dag}$ creates an electron at site
$i$ with $z$-component of spin $\sigma=\uparrow,\downarrow$,
$n_i=\sum_{\sigma} c_{i,\sigma}^{\dag}c_{i,\sigma}$, and ${\bf S}_i$
are electron spin operators.  The projector
${\cal P}=\prod_i(1-n_{i,\uparrow}n_{i,\downarrow})$ projects out states
with doubly occupied sites.  By taking the $V \rightarrow 0$ limit,
one recovers the original \mbox{$t$-$J$}\ model.  Secondly, we will
study a model which includes next-nearest-neighbor repulsions:
\begin{equation} \label{eq:htjvv}
        H_{t-J-V-V'}=H_{t-J-V}+ V'\sum_{i}n_in_{i+2}\;.
\end{equation}
The last modification is a model with a bare Coulomb repulsion
and a spatial decay proportional
to $1/r$ \cite{barnes}.  For a finite lattice, we will use the following
Hamiltonian,
\begin{eqnarray} \label{eq:htjvr}
     H_{V/r}&=&-t\sum_{i,\sigma}
     \left({\cal P}c_{i,\sigma}^{\dag}c_{i+1,\sigma}{\cal P}+h.c.\right)
     \nonumber \\
     &&+J
     \sum_{i}\left({\bf S}_i \cdot {\bf S}_{i+1}
        - {1\over4} n_i n_{i+1}\right)
     + {V_L \over 2} \sum_{i,j}{(n_in_j -\rho ^2)\over d_{ij}} \;,
\end{eqnarray}
where $\rho \equiv (\sum_i n_i )/L$ is the electron concentration, and
the ``distance" $d_{ij}$ between the sites $i$ and $j$ is defined as
\begin{equation}
   d_{ij}={L\over2\pi} \sin\left({2\pi\over L}|i-j|\right)\; ,
\end{equation}
on our finite lattice of $L$ sites.  In the $L \rightarrow \infty$ limit,
$d_{ij} \rightarrow |i-j|$.  The last term in the Hamiltonian
is a constant and gives the contribution of a uniform positive-charge
background, which keeps the ground state energy per site finite in the
limit $L \rightarrow \infty$.

%
%
\section{Numerical methods}

We use both exact diagonalization and QMC methods to
investigate the ground state properties of the models described in the
previous section. Exact
diagonalization gives results with high accuracy for small
lattices. It also allows the calculation of dynamical properties in
real time. QMC can be used to investigate static properties of
larger lattices.
For the exact diagonalization method,
the Lanczos algorithm \cite{lanczs} is
used to obtain the energy eigenvalues and the eigenvectors for the
ground state as well as for the first few excited states.
The relative error of the eigenvalues, and the residue of the eigenvector are
both less than $10^{-9}$ for lattices of up to 20 sites.
In order to carry out a systematic finite size analysis, the
boundary conditions must be chosen carefully. One choice is to select
boundary conditions which keep all one particle orbitals either fully
occupied or empty in the noninteracting case. We will call this choice
closed shell boundary conditions (CSBC). In practice, we use
periodic boundary conditions (PBC) for systems with $N=4n+2$ particles
($n$ is an integer) and antiperiodic boundary conditions (APBC)
for $N=4n$ particles. The opposite choice of antiperiodic boundary
conditions for systems with $N=4n+2$ particles and periodic boundary
conditions for $N=4n$ particles will be called open shell boundary
conditions (OSBC). The ground state of the 1D \mbox{$t$-$J$}\
model is always a spin singlet with CSBC, but it changes the spin quantum
number depending on the parameters for OSBC.
In most cases we use CSBC.
In cases where a more careful analysis is necessary to study finite
size effects, we use both CSBC and OSBC.

By QMC methods considerably larger systems can be studied at low
temperatures. Here, we use the world line algorithm \cite{Suzuki} with
a four-site cluster decomposition \cite{cluster}.
Since there is no negative sign problem for the 1D
\mbox{$t$-$J$}\ model \cite{assaad}, simulations could be performed on
lattices with up to $L=96$ sites at inverse temperatures up to
$\beta t=64$. The systematic error of order ${\rm O}(\Delta\tau^2)$
due to the finite Trotter time
step $\Delta\tau$ is controlled by  extrapolating the results
obtained at
$\Delta\tau t=0.25$ and $\Delta\tau t=0.5$. The usual zero winding
number boundary conditions were used.  For more details on the
algorithm we refer to Refs. \cite{assaad,Suzuki,cluster}.

To study the properties of the \mbox{$t$-$J$-$V$}\ model
we first calculate the spin gap by exact diagonalization and QMC.
Next we calculate the charge and spin structure factors by both methods.
Furthermore we measure pairing correlations by exact diagonalization.
The charge and spin structure factors are defined as the Fourier
transform of the correlation functions in real space:
\begin{eqnarray}
   S_{charge}(q)&=&{1\over L}\sum_{j,m}^Le^{iq(j-m)}
     \langle (n_{j,\uparrow}+n_{j,\downarrow})
	   (n_{m,\uparrow}+n_{m,\downarrow}) \rangle , \\
   S_{spin}(q)&=&{1\over L}\sum_{j,m}^Le^{iq(j-m)}
     \langle (n_{j,\uparrow}-n_{j,\downarrow})
	    (n_{m,\uparrow}-n_{m,\downarrow}) \rangle \nonumber\\
    &=&{4\over L}\sum_{j,m}^L e^{iq(j-m)}
       \langle S_j^z S_m^z \rangle.
\end{eqnarray}
The singlet pairing correlations are defined as
$\langle P^{\dag}(r) P(0)\rangle$,
where
\begin{equation} \label{eq:delta}
   P^{\dag}(r)
     = {1\over \sqrt{2}}
       \left( c^{\dag}_{r,\uparrow}c^{\dag}_{r+1,\downarrow}
	     -c^{\dag}_{r,\downarrow}c^{\dag}_{r+1,\uparrow}
       \right) \;
\end{equation}
is the creation operator of a nearest-neighbor electron singlet pair.
The superconducting structure factor is defined as its Fourier transform, i.e.
\begin{equation} \label{eq:chi}
   S_{pair}(q) ={1\over L} \sum_{j,m}^L  e^{iq(j-m)}
	\langle P^{\dag}(j) P(m) \rangle \; .
\end{equation}

%
%
\section{Phase separation}

In the \mbox{$t$-$J$}\ model it is well known that
phase separation occurs for
large values of $J/t$. At low hole doping this effect arises
in order to minimize the number of broken antiferromagnetic bonds
in the system. For low electron doping, the $J$-term is an
explicitly attractive term for electrons. For large values of $J$
this attraction overcomes
the repulsion due to the kinetic term and the system is separated
into a particle-rich phase and a sea of holes.  Close to the phase
separation boundary the system is still homogeneous but the attraction
already produces bound states of holes or electrons, depending on the
doping. Precisely this effect led to recent studies \cite{dagotto,dagotto2}
of that region in both 1D and 2D to search for
indications of superconductivity in this type of models. Superconductivity
will be investigated in detail in Sec. V.

In this section, we will determine the boundary of the phase separation
region by exact diagonalization techniques. This boundary is defined by
those points in parameter space
where the inverse compressibility $\kappa^{-1}$ vanishes.
On a lattice with $N$ particles and $L$ sites, $\kappa^{-1}$
can be calculated as
\begin{equation}
  \kappa^{-1}={N^2\over L}
     \left({E(N+2;L)+E(N-2;L)-2E(N;L)\over 4}\right) \;,
\end{equation}
where $E(N;L)$ is the ground state energy of the finite system with $N$
particles on $L$ sites. Equation (9) is simply a discrete version of the
second
derivative of the energy with respect to the number of particles.
The phase separation boundary can also
be estimated using QMC methods. Here, phase separation is characterized
by a divergence of the long-wavelength charge fluctuations
$S_{charge}(q=2\pi/L)$ when the system size $L$ is increased (for
details we refer the reader to
Ref.~\cite{assaad}). The results obtained by both methods agree
very well. Comparing results for different system
sizes $L$, we estimate the error on this boundary to be of the
order of $\Delta J\sim 0.1t$.

Figure \ref{kappa} shows the inverse compressibility for the
\mbox{$t$-$J$}, \mbox{$t$-$J$-$V$}, and
\mbox{$t$-$J$-$V$-$V'$}\ models at quarter band-filling.
It can clearly be seen that, as expected, the repulsive $V$-term
pushes the phase separation boundary to larger couplings $J/t$.
In the phase diagrams (Figs.~\ref{phasediag}
and \ref{phasediagrj}), this behavior is clearly seen in the $J$-dependence
of the phase separation line (thick line in the figures).
The next-nearest-neighbor repulsion $V'$ shifts phase separation
into even larger values of $J$.

The case of a long-range repulsion of $1/r$-form is different.
On a finite size lattice, a large enough $J>J_c$ will result in
phase separation but the critical value $J_c$ diverges with the system
size. Thus, in an infinite lattice no phase separation
will occur. This is easily understood since the contribution of the long-range
repulsion to the energy diverges in the phase separated state. When
$J$ is increased the system does not phase separate but rather forms a charge
density wave (CDW) consisting of larger and larger clusters
of antiferromagnetically aligned spins.

%
%
\section{Spin gap}

Several well-studied models of interacting electrons present a nonzero
spin gap. For example, the
extended Hubbard model for $V>U/2$
has a gap in the spin excitation spectrum \cite{solyom}.
The attractive-$U$ Hubbard model in 2D has both a superconducting
and a spin-gap at all fillings \cite{adriana}.
The Luther-Emery model is an example of an exactly solvable model
with a spin gap \cite{lutheremery}.
Also spin models with ground states made only of local spin singlet pairs,
e.g. the AF
Heisenberg chain with a frustrating next-nearest-neighbor interaction
\cite{madjumdar}, have a nonzero spin gap.
Doping holes into such a chain leads to the
\mbox{$t$-$J$-$J'$}\ model with charge carriers
which also exhibits a spin gap near half
filling \cite{ogata3}.
However, studies of the
\mbox{$t$-$J$}\ model in 1D
at quarter band-filling \cite{ogata} did not show indications of such a
gap. This has to be contrasted against the recent analysis of the
\mbox{$t$-$J$-$V$}\ model by Dagotto and Riera where at large and intermediate
values of $V/t$, a nonzero spin gap was observed. In this section, we
clarify this situation by an analysis of the spin gap in the
\mbox{$t$-$J$-$V$}\ model at all values of the coupling $V/t$ using Lanczos
and QMC methods.



First, let us consider
the classical limit $t=0$. In this case the ground state can be obtained
exactly following, for example, the procedure of Ref.\cite{dagotto}.
Both the \mbox{$t$-$J$-$V$}\ and
\mbox{$t$-$J$-$V$-$V'$}\ models have a spin gap for
$V-2V'<J<(V+2V')/(2\log2-1)$ at quarter band-filling.
In this parameter region
the ground state consists of nearest-neighbor singlet pairs separated by
two holes ($2k_F$ CDW). Other configurations are also possible in this
regime since a pair of electrons in the previous state can be moved one
lattice
spacing to the right or the left without additional cost in energy.
A similar analysis
can be carried out in higher dimensions leading also to the presence of
a spin gap \cite{dagotto,kivelson}.  Therefore the spin excitations have a gap
$\sim J$.
The problem is whether this spin gap will survive for finite $t$. To
study this issue
we calculated the spin gap using exact diagonalization methods on lattices
with up to $L=20$ sites at quarter band-filling ($\rho=1/2$).
The spin gap on a finite
lattice, $\Delta(L)$, is evaluated directly as the difference between
the energies of the lowest lying spin singlet and triplet states.
The spin gap $\Delta$ in infinite systems is then obtained by
extrapolating $\Delta(L)$ to the bulk. As a scaling function we have used
the following form (for fixed fillings $\rho=N/L$):
\begin{equation} \label{fit}
   \Delta^{BC}(L)
    =E_0^{BC}(L)-E_1^{BC}(L)
    =\Delta+{a_1^{BC}\over L}+{a_2^{BC}\over L^2}\; ,
\end{equation}
where $E_0^{BC}(L)$ and $E_1^{BC}(L)$ are the lowest eigenenergies
in the spin singlet and triplet subspaces, respectively. The superscript
$BC$ denotes the boundary conditions. Both CSBC and OSBC were used.
We have observed that
including a $1/L^3$ term changes $\Delta$ by only a few percent, and
thus our results seem stable. We have tested this extrapolation using the
 \mbox{$t$-$J$}\ model,
where no spin gap was reported at quarter band-filling.
In the case of a finite spin gap we
expect that asymptotically for large $L$ the spin gap follows an
exponential scaling. The results of
QMC for systems with $L=48$ and $L=96$ sites indicate such a behavior,
but for the small system sizes that can be investigated using exact
diagonalization we are not yet in this exponential regime and the
fitting function Eq.(\ref{fit}) is the best.
In the QMC simulations the spin gap is calculated as the difference between
the energies of the $M_z=0$ and $M_z=1$ subspace. ($M_z$ is the
$z$-component of the total magnetization).
As shown in Figure~\ref{gapfit},
the scaling function Eq.(\ref{fit}) somewhat
underestimates $\Delta$, as expected from the correct asymptotic exponential
behavior. Therefore,
the spin gap $\Delta$ calculated following our procedure should be
considered as a $lower$ bound for the actual value of the gap.

In Figure~\ref{gap} the spin gap for the quarter-filled
\mbox{$t$-$J$-$V$}\ model is shown. The parameters are chosen
along the line $J=3t+2V$, which is inside the superconducting region
close to phase separation. The spin gap $\Delta$
increases with $J$ and $V$. In the phase diagram shown
in Fig.~\ref{phasediag}, we plot the contour lines for several
values of constant gap $\Delta$.
A prominent feature is that the spin gap
region expands with increasing $V$, which is consistent with the
result in the classical limit $t=0$.
As mentioned above,
the effect of the $V'$ term
on the spin gap is also estimated qualitatively by considering
the limit $t\rightarrow0$. The $V'$ term $expands$ the spin gap region in the
strong coupling regime.

Figure 5 shows that the spin gap is nonzero in the bulk limit for all
the values of $V/t$ that we have analyzed, starting at $V/t = 0.5$.
Then, it may occur that for a certain range of $J/t$-values $V/t=0$ is a
singular point and that a small perturbation away from it,  opens a
gap immediately.
 Such a behavior was recently suggested \cite{dagotto}, but by no
means proved, based on a mapping
of the large $V/t$ results into the attractive Hubbard model. We know that
the last model opens a spin gap as soon as the interactions are turned on.
Of course, we cannot exclude the possibility that for all values of $J/t$
there is a
``critical'' value of $V/t$ larger than $0$, where the gap opens.
We expect that the spin gap is sensitive to the value of $J/t$.  The
investigation of this behavior certainly deserves more work.

%
%
\section{Superconductivity}

In this section we will study superconducting correlations,
and discuss their
relation to the phase separation and spin gap regions. As discussed in
Sec. I, our main concern is whether intermediate and
long-ranged Coulomb interactions can
enhance superconductivity by suppressing  phase separation,
which destroys superconductivity otherwise.
For this purpose,
we have calculated several quantities to investigate superconducting
correlations in the ground state.
First, we have measured the absolute value of the superconducting structure
factor $S_{pair}(q=0)$. Then, we study the real-space pairing correlations
$\langle P ^{\dag}(r)P(0) \rangle$ as a function of distance. Finally, we
will investigate the width of the superconducting region by analyzing
the correlation exponents in Sec. VII.

In the \mbox{$t$-$J$-$V $}\ model, it was shown by Dagotto and
Riera \cite{dagotto} that increasing the value of $V/t$ from zero,
the nearest-neighbor density-density repulsion
enhances both the peak value of $S_{pair}(q=0)$, and the
pairing correlations at large distances
compared to the \mbox{$t$-$J$}\ model. Furthermore, a similar qualitative
behavior was observed in the more realistic 2D case
\cite{dagotto,dagotto2}.
For larger values of $V/t$, the pairing correlations are eventually
suppressed and
they decay to zero when $V/t \rightarrow \infty$, due to the lack of mobility
of the pairs \cite{dagotto}. In
these previous studies, it was observed that the
inclusion of the repulsive term shifts not only the phase separation
boundary to larger values of $J/t$, but the superconducting region
is shifted as well, always existing as a narrow strip close to phase
separation.

Since the nearest-neighbor term enhances
superconductivity \cite{dagotto},
let us consider the effect of a
next-nearest-neighbor repulsion term $V'$ (Hamiltonian
Eq.(\ref{eq:htjvv})), which further suppresses phase separation.
Figures \ref{suprv1} and \ref{plot} show the uniform component
of the superconducting structure factor $S_{pair}(q=0)$ for $V/t=1$
at quarter band-filling ($\rho=1/2$) and for $\rho=2/3$, respectively.
A small repulsion
($V'/V=0.25$ and $V'/V=0.5$) enhances the peak value of the
superconducting structure factor for both dopings.
The maximum is around $V'/V =0.5$. At large values of $V'>V/2$
superconductivity is reduced, and actually
at $V'/t=2$ it is strongly suppressed. For larger values of $V/t$, the
effect of $V'/t$ is less important. Actually, for $V/t \sim 3$ or
larger, we observed that the $V'$-term suppresses superconductivity
as soon as it is turned on.
However, these values may be unphysically large in the real materials.
As in the \mbox{$t$-$J$}\ model
the superconducting structure factor is largest in a region near
the phase separation boundary. The suppression of superconductivity
with large $V'$ is due to a competition with CDW correlations, as will
be shown in the next section.

As emphasized in a recent paper \cite{dagotto}, the $q=0$ component of the
superconducting structure factor,
$S_{pair}(q=0)$, contains both the short and long distance
correlations. Thus, it is very important to study the pairing
correlations in real space to observe its asymptotic behavior.
For the \mbox{$t$-$J$-$V$}\ model ($V'=0$), and
the \mbox{$t$-$J$-$V$-$V'$}\
model at $V'=V/2=t/2$, we have calculated
$\langle P ^{\dag}(r)P(0) \rangle$ for quarterband-filling on lattices
of $L=12,16$ and $20$ sites, and, additionally, for a filling of
$\rho=2/3$ on lattices of $L=12,15,18$. In Fig.~\ref{plots} several
typical results are plotted for values of $J/t$ at the maximum of the
superconducting structure factor.
They show an increase in the long-range pairing
correlations with $V'$, as suggested by Fig.7.
The features become more prominent with
an increase of the system size. From this evidence we conclude that
our results are valid for the infinite system and the next-nearest-neighbor
interaction indeed further increases the long-range pairing correlations.
The behavior observed in the susceptibility is thus indicative of long
distance properties of the ground state.

As mentioned above, the system with a $1/r$ long-range repulsion does not
undergo phase separation in the thermodynamic limit. However, the long-range
repulsion also suppresses the superconducting pairing correlations (see
Figs.~\ref{plot} and \ref{plots}) at large distances. For this particular
example, note that the study of the large distance behavior of the
correlations is crucial. From Fig.~\ref{plot} it would have been concluded that
$1/r$ interactions also enhanced superconductivity.
However, this is a short distance effect. Actually, in Fig.~\ref{plots} for
distances
smaller than three lattice spacings, correlations for different models
are all similar, while
only at large distances it can be observed that the correlations for the
$1/r$ interaction are strongly suppressed.

To summarize the results of this section,
we have observed that a short-range Coulomb repulsion suppresses
phase separation, and enhances both the superconducting structure factor
and the long-distance pairing correlations. For the particular case
of $V'=0$ this is in agreement with previous results \cite{dagotto}.
On the other hand, the long-range $1/r$ repulsion does not enhance
pairing correlations in the ground state, in spite of the fact that it
suppresses phase separation. The reason is that $two$ effects are in
competition against superconductivity: one is phase separation, and the
other is CDW order. In other words, in a region of electron pairs, as
that found in the $t=0$ limit, we can have superconducting $or$ CDW
correlations in the ground state once the hopping $t$
becomes nonzero. This
can be clearly seen in the attractive Hubbard model where at
half-filling,
both types of orders are exactly degenerate.
Away from that special point we expect the degeneracy to be lifted. While
phase separation is pushed further away by the long-range repulsion
both CDW and superconducting correlations are enhanced, but
the CDW correlation are dominant.

%
%
\section{Charge and spin structures}

In this section we will examine the competition between phase separation,
superconductivity, CDW and spin density wave (SDW) order in the ground
state of the several models under consideration.
We have calculated
the correlation functions and structure factors of the \mbox{$t$-$J$-$V$}\
model Eq.(\ref{eq:htjv}) on large systems with up to $L=96$ by using
QMC techniques at $\rho=1/2$ and $V/t=1$. For the long-range $1/r$ model
Eq.(\ref{eq:htjvr}), we have carried out exact diagonalization calculations
for systems of up to $L=20$ sites, since the world line algorithm
of the QMC method requires the interactions to be local.

At small values of $J/t$ the \mbox{$t$-$J$-$V$}\ model
shows similar qualitative behavior of the charge and spin structures
as the \mbox{$t$-$J$}\ model.
The system is characterized by $4k_F$ CDW correlations showing
a power law decay like in the \mbox{$t$-$J$}\ and Hubbard models.
The spin degeneracy of the case $J=0$ is lifted by an infinitesimal
$J$, leading to dominating $2k_F$ SDW fluctuations. Consequently
the charge and spin structure factors present a cusp at $q=4k_F$
and $q=2k_F$, respectively ($J/t=0.5$ in Fig.~\ref{corr1}a and b).

Let us now increase $J/t$, while keeping $V/t$ fixed. At larger values
of $J/t$ the particles form nearest-neighbor singlet
pairs, as was discussed in previous sections and Refs.
\cite{dagotto,kivelson}.
This effect suppresses the $4k_F$ charge and $2k_F$ spin fluctuations,
while enhancing $2k_F$ charge fluctuations. In the structure factors
this effect can be seen by a decrease in the $2k_F$ spin singularity
and the development of a $2k_F$ charge structure ($J/t=4$ in
Figs.~\ref{corr1}a and b). While the Coulomb repulsion enhances
the formation of nearest-neighbor singlet pairs, the CDW correlations
still dominate the pairing correlations.  The real-space spin
correlations show a strong AF nearest-neighbor correlation.  The
amplitude of the longer-range correlations is, on the other hand,
very small. This makes it hard to distinguish
the TLL region, where
the spin correlations show a power-law decay, from the spin gap
region where they show an exponential decay. Therefore it is necessary
to combine the CFT and numerical calculations to obtain the correlation
exponents as will be done in the next section.

At even larger values of $J/t$, the system is dominated by the
superconducting correlations.  As a typical example, charge and spin
correlation functions are shown for $J/t=4.75$ in Fig.~\ref{corr1}a and b.
There the particles tend to form nearest-neighbor singlet pairs
and the spin excitations have a gap.  This is reflected by the
spin structure factor which is similar to that of a gas of
nearest-neighbor singlet pairs.

Finally, at $J/t>5$ phase separation occurs. The inverse compressibility
becomes zero and the $q=2\pi/L$ component of the charge structure factors
increases strongly and diverges with increasing system size.

If the long-range interactions of $1/r$ form are used, the CDW correlations
are dominant for the whole-range of couplings $J/t$ as can
be seen from the charge structure factor (Figs.~\ref{corr2}a and b).
At small $J/t$, we can again
see a $4k_F$ cusp in the charge structure factor, and a cusp at
$2k_F$ in the spin structure factor ($J/t=2$ in Fig.~\ref{corr2}a and c).
Increasing $J/t$, a $2k_F$ peak in $S_{charge}(q)$ develops,
while the spin structure has a maximum at $q=\pi$ and looks similar
to that of a gas of nearest-neighbor singlet pairs ($J/t=5$ in
Fig.~\ref{corr2}a,c).  At larger $J/t$ there is no phase separation
but, as discussed before, the particles form larger and larger
clusters of AF spin chains. This can be seen in the charge structure
factor as the shift of the peak towards smaller $q$.  The spin structure
factor resembles that of an AF Heisenberg chain with a peak at
$q=\pi$. In the long-range model the CDW correlations always dominate
the superconducting correlations.

%
%
\section{Correlation exponents}

In this section we will calculate the correlation exponents of the
\mbox{$t$-$J$-$V$}\ model. The correlation exponents can be used
to decide which correlations dominate the long-range behavior.
It is in general hard to determine the correlation exponents directly
from the numerical calculations of the correlation functions.
However, in many 1D Fermion systems, by combining conformal field theory
(CFT) with
numerical simulations, one can determine the correlation exponents
from thermodynamic quantities which can be calculated more easily
and accurately than the long-range correlations \cite{solyom,haldane}.

Depending on the coupling constants two main regimes can be distinguished.
In the TLL regime both the charge and spin excitations are gapless
and the correlation functions show a power-law decay. The exponents
can be described by a single dimensionless exponent $K_\rho$
\cite{solyom,haldane}:
\begin{eqnarray} \label{crit}
   \langle n(r)n(0)\rangle & & \sim A_{0}r^{-2} + A_{1} \cos(2k_Fr) \;
	r^{-(1+K_{\rho})} + A_{2} \cos(4k_Fr) \;
	r^{-4 K_{\rho}}, \nonumber \\
   \langle S_{z}(r)S_{z}(0)\rangle & & \sim B_{0} \;
	r^{-2} + B_{1} \cos(2k_Fr) \; r^{-(1+K_{\rho}) }, \\
   \langle P^{\dagger}(r)P(0)\rangle & & \sim C_{0}
	r^{-(1 + \frac{1}{K _{\rho}} )} + C_{1} \cos(2k_Fr) \;
	r^{-(K_{\rho} + \frac{1}{K_{\rho}})}. \nonumber
\end{eqnarray}
Logarithmic corrections have been omitted in these formulas.
Models belonging to the TL regime include the
\mbox{$t$-$J$}\ model before phase separation and the repulsive
Hubbard model, as well as many generalizations of these models
without a spin gap.
The other regime is the spin gap phase, typically represented
by the Luther-Emery model.  This phase has a finite  gap in the
spin excitation spectrum but gapless charge excitations.
Here the spin correlations have an exponential decay, and the other
correlations again present a power law decay:
\begin{eqnarray} \label{critsg}
   \langle n(r)n(0)\rangle & & \sim A_{0}r^{-2} + A_{1} \cos(2k_Fr) \;
       r^{-K_{\rho}} + A_{2} \cos(4k_Fr) \; r^{-4 K_{\rho}}, \nonumber \\
   \langle P^{\dagger}(r)P(0)\rangle & & \sim
       C_{0} r^{-\frac{1}{K _{\rho}}} .
\end{eqnarray}
The extended Hubbard model includes both the TLL and spin gap
phases. It is apparent that the superconducting pairing correlations
are dominant for $K_\rho>1$ in both phases.

Because the \mbox{$t$-$J$-$V$}\ model shows qualitatively similar
behavior in the correlation functions as the \mbox{$t$-$J$}\
model, we expect it to belong to the same universality class.
This is further confirmed by calculation of their
central charges. The CFT predicts the following finite
size corrections of the ground state energy for a TLL \cite{haldane}:
\begin{equation}
\label{eq:cc}
   E(L)/L=\epsilon-{\pi\over6}(v_c+v_s){c\over L^2}\;,
\end{equation}
with the central charge $c=1$. Here, $E(L)$ is the ground state
energy of the finite system with $L$ sites and $\epsilon$ is the
energy per site in the infinite system.  The parameters, $v_c$
and $v_s$, are the charge and spin velocities, respectively. They
can be calculated as the derivative of the charge and spin excitation
energies with respect to total momentum. We  calculated them by using
\begin{eqnarray}
v_c&=&\left[E_0(q=2\pi/L)-E_0(q=0)\right]/{2\pi\over L} \\
v_s&=&\left[E_1(q=2\pi/L)-E_0(q=0)\right]/{2\pi\over L},
\end{eqnarray}
where $E_0(q)$ and $E_1(q)$ are the lowest energy eigenvalues in the spin
singlet, respectively triplet, subspace of total momentum $q$. The central
charge
was then obtained by fitting the ground state energies according to
Eq. (\ref{eq:cc}). Our calculations
give $c\sim1.04$, which is in agreement with $c=1$ within the
finite size errors in the velocities of the order of a few percent.
This value of $c$ is nearly $J$ and $V$ independent. The deviations from
unity are always smaller than the estimated error in the velocities which
are about $10\%$.

The correlation exponent $K_\rho$ can now be calculated from the relation
\cite{haldane}
\begin{equation}
   K_\rho=\pi v_c\rho^2\kappa/2\;,
\end{equation}
with $\rho$ being the particle density, and $\kappa$ the compressibility.
In Fig.~\ref{phasediag} we show contour lines of constant $K_\rho$ in the
$J$-$V$ plane at quarter band-filling. In addition, Fig.~\ref{phasediagrj}
shows the phase diagram of the \mbox{$t$-$J$-$V$}\ model at $V/t=1$
for all fillings.

Several regions can be distinguished in the phase diagram.  For
large values of $J$ the system is phase separated.  As a precursor
to phase separation there is a region of width $\Delta J \sim
t$ where the correlation exponent $K_\rho>1$, and thus the superconducting
correlations correspondingly dominate the long-range behavior.
At smaller values of $J/t$, $K_\rho < 1$ and, therefore,  the
CDW correlations are dominant. Another important line is the boundary
of the spin gap region (the line $\Delta=0$ in the phase diagram).
Although the lines of constant $K_\rho$ seem to be continuous
at that boundary, a large change in the long-range correlations
occurs there. A direct consequence is, of course, that the spin
correlation functions change their space-time dependence from
a power law in the TLL phase to an exponential form in the spin
gap phase.  A more noticeable effect is the jump in the exponents of charge
and superconducting correlations.  The charge exponent jumps from
$1+K_\rho$ to $K_\rho$, while the superconducting exponent
$1+1/K_\rho \rightarrow 1/K_\rho$ (compare Eqs.~(\ref{crit}) and
(\ref{critsg})).  As a consequence, the exponent of the most dominant
correlations is always greater than or at least equal to unity
in the spin gap phase.  This means that the corresponding structure
factor is divergent.  On the other hand, in the TLL phase, even
the most dominant exponent is less than unity, leading to a cusp
in the structure factors instead of a $divergent$ peak.
This implies a strong increase in the superconducting correlations
in the spin gap region due to the repulsive interactions.

%
%
\section{Conclusions}

We have examined in  detail the precursor region to phase
separation for the 1D \mbox{$t$-$J$}\ model including repulsive
density-density
interactions.  These repulsive interactions, as expected, move
the phase separation boundary to larger value of $J/t$ but at
the same time open up a wider precursor region where strengthened
superconducting correlations appear.  Our numerical calculations
confirm earlier reports of superconductivity in the precursor
region and show clear evidence for the existence of a spin gap
and an enhanced range of superconducting correlations for short-range
repulsions \cite{dagotto,dagotto2}.  However, we find that
the form of the repulsive interaction
strongly influences the precursor region, and in certain cases, such
as a long-range $1/r$-interaction, it favors a CDW state over
a superconducting state.

What is
the relevance of these results to the \mbox{high-$T_c$}\ superconductors?
At first sight, it might seem that our results have no relation with
the cuprates since they have been obtained in 1D, at large $J/t$ and
mainly at dopings $\rho = 1/2$ and $2/3$. However, note two
important issues: first, recent work has shown that regarding
superconductivity there is no drastic qualitative difference between 1D and 2D
\cite{dagotto,dagotto2}. Although there is no
possibility of real long-range order in 1D,
both in 1D and 2D superconducting correlations
seem dominant in the same region, namely close to phase separation.
Second, it has been observed that numerically the signals of superconductivity
are maximized near quarter filling in both cases.

Then, it is possible
to make some speculations about 2D models based on the results
of the 1D systems. For example, note that in Fig. 3
the region where superconductivity dominates exists at dopings as
close to half-filling as $\rho \sim 0.875$.
However, the numerically calculated amplitude of the superconducting
correlation decreases when $\rho$ changes
from $1/2$ to $2/3$, as shown in Figs. 7 and 8. This result implies that
superconductivity may exist near half-filling but is difficult to
observe numerically through pairing-pairing correlation functions,
mainly because the number of pairs contributing to superconductivity decreases
to zero when $\rho \rightarrow 1$.
Certainly, it could be that we have a similar situation in 2D, i.e.
that the strip of superconductivity ($K_\rho>1$) at
quarter-filling \cite{dagotto2} reaches the neighborhood of
half-filling, with a small order parameter hard to detect numerically.
In the 2D \mbox{$t$-$J$} model phase separation near half-filling
starts at a coupling close to $J/t \sim 1$, and thus the region of
superconductivity may exist in the realistic region of $J/t \sim 0.3
-0.4$ and low doping. We believe that it is very important to test
these speculations.

\acknowledgments

The quantum Monte Carlo calculations were done
on the Cray Y/MP-464 of ETH Z\"urich. The exact diagonalization studies
were performed on the NEC SX-3/22 of the Centro Svizzero di Calcolo
Scientifico CSCS Manno, and also at the CRAY-2 of the National
Center for Supercomputing Applications, Urbana, IL, USA.
The work was supported by the Swiss National
Science Foundation under grant number SNF 21-27894.89 and by an internal
grant of ETH Z\"urich. We wish to thank F.F. Assaad, M. Imada,
M. Luchini, A. Moreo, M. Ogata, and D. W\"urtz for helpful discussions.


%
%

%
%
%
\figure{
Inverse compressibility for the \mbox{$t$-$J$}\ model,
the \mbox{$t$-$J$-$V$}\ model (at $V/t=1$) and the
\mbox{$t$-$J$-$V$-$V'$}\ model (at $V/t=1$, $V'/t=0.5$) at quarter
band filling on a chain with $L=16$.
The phase separation boundary ($\kappa^{-1}=0$) is shifted
towards larger $J/t$ with inclusion of Coulomb repulsions.
\label{kappa}
}
%
%
\figure{
Phase diagram of the 1D \mbox{$t$-$J$-$V$}\ model
at quarter filling.
The thick line denotes the phase separation boundary.
Contour lines of constant $K_\rho$ (solid lines) and constant
spin gap $\Delta$ (dashed lines) are shown.
The error of $\Delta=0$ line is estimated about $\Delta J\sim0.5t$,
while it is
about $\Delta J\sim0.1t$ for the other $\Delta$ and $K_\rho$ lines.
 These contour lines were obtained from interpolating
results on a grid with a spacing of $0.25t$ in the $J$ and $V$ direction.
\label{phasediag}
}
%
%
\figure{
Phase diagram of the 1D \mbox{$t$-$J$-$V$}\ model in
the $\rho$-$J$ plane for $V/t=1$.  $L=16$.
The solid line is the phase separation boundary, and the others
are contour lines of constant $K_\rho$.
These contour lines were obtained from interpolating
results on a grid with a spacing of $0.25t$ in the $J$ direction for systems
with an even number of particles and CSBC. The errors on the $K_\rho$-lines
are estimated to be smaller than $\Delta J\sim 0.1t$
\label{phasediagrj}
}
%
%
\figure{
Finite size scaling of the spin gap in the \mbox{$t$-$J$-$V$}\
model for $J/t=9$ and $V/t=3$ at $\rho=1/2$.
Data for up to $L=20$ are calculated by exact diagonalization with
CSBC and OSBC.  Data for $L=16,24,48$ and $96$ are calculated by QMC
at an inverse temperature of $\beta t=24$.
Included are the extrapolations
to the $L\rightarrow\infty$ limit using polynomials of second,
respectively third, order in $1/L$.
\label{gapfit}
}
%
%
\figure{
The extrapolated spin gap along the line $J=3t+2V$
in the \mbox{$t$-$J$-$V$}\ model at $\rho=1/2$.
\label{gap}
}
%
%
\figure{
Superconducting structure factor for the \mbox{$t$-$J$-$V$-$V'$}\
model for $V/t=1$ and various values of $V'$
at quarter band-filling.  L=16 with CSBC.
\label{suprv1}
}
%
%
\figure{
Superconducting structure factors of the \mbox{$t$-$J$-$V$},
\mbox{$t$-$J$-$V$-$V'$}, and the long-range repulsion models.
(a) 10 particles on 20 sites ($\rho=1/2$);
(b) 12 particles on 18 sites ($\rho=2/3$).
Both CSBC and OSBC are used. The small peak around $J/t=8$
for the long-range model is due to finite size effects
and vanishes with increasing the system size. The jump seen
for OSBC at $J/t\sim4$ is due to a level crossing of the ground state.
At small values of $J/t$, the ground state is spin triplet for OSBC.
With the opening of the spin gap, the singlet spin state
becomes the ground state.
\label{plot}
}
%
%
\figure{
(a), (b)
Pairing correlations for the same models as in Fig.~7. at the $J$-values
where the superconducting structure factor has its maximum for that model.
Note that in the finite system this peak, while being close to the phase
separation boundary, is actually at $J$ values where the infinite
system is phase separated. In the infinite system the peak is located exactly
at the phase boundary.
(c) Pairing correlations for $L=20$ and $\rho=1/2$ with CSBC for three
different $J$-values: $J/t=3$, where CDW correlations are dominant; $J/t=5$
where superconductivity is dominant; and at $J/t=7$, which is in the phase
separated region. The increase of the pairing correlation for $J/t=7$ at
$r=10$ is a finite size effect which vanishes when larger lattices are
considered.
\label{plots}
}
%
%
\figure{
Monte Carlo results of (a) charge and (b) spin structure factors
in the \mbox{$t$-$J$-$V$}\ model at $V/t=1$ for different values
of $J/t$.
The lattice size was $L=96$ sites with $N=48$
particles, the inverse temperature $\beta t=24$ or $\beta t=64$.
The imaginary time step $\Delta t=0.25$ was chosen small enough
to see the properties of the ground state.
\label{corr1}
}
%
%
\figure{
Charge and spin structure factors for the long-range model obtained by exact
diagonalization.  $L=20$ and $N=10$ with CSBC.  (a) Charge structure factors
for small values of $J/t$.  (b) Charge structure factors for large values
of $J/t$. (c) Spin structure factors for the entire coupling range.
\label{corr2}
}

\end{document}